\documentstyle[preprint,aps]{revtex}
\begin{document}

\title{Algebraic Self-Similar Renormalization in Theory of Critical Phenomena.}
\author{S.Gluzman and V.I. Yukalov} 
\address {International Centre of Condensed Matter Physics, University of\\
Brasilia, CP 04513, Brasilia, DF 70919-970, Brazil} 
\maketitle
\begin{abstract}
We consider the method of self-similar renormalization for calculating
critical temperatures and critical indices. A new optimized variant of the
method for an effective summation of asymptotic series is suggested and
illustrated by several different examples. The advantage of the method is in
combining simplicity with high accuracy.
\end{abstract}
\section{Introduction}

In the theory of critical phenomena one usually obtains critical
temperatures and critical indices as expansions in powers of some parameters
which, as a rule, are not small. For instance, in the Wilson $\epsilon -$%
expansion [1] one has $\epsilon =1.$ In the field-theoretical approach [2]
the expansion is in powers of the renormalized coupling constant $g\cong
1.4.\ \ $Such expansions, as is well known, are asymptotic and lead to
reasonable results only in the low orders. The direct use of higher orders
makes the results only worse [2,3]. When a {\it number of terms} in a
divergent series is known one may invoke resummation techniques, such as Pade
and P\'ade-Borel ones. While the knowledge of only a {\it few first terms}
does not permit to use these techniques. Thus, one always confronts the
problem of how to improve the results of a divergent series having only a
few terms.

In the present paper we suggest a solution of this problem advancing a
method which has the following peculiarities: (i) \ It permits to accomplish
an effective summation of a divergent series consisting of just a couple of
terms, when no other method is applicable; (ii) It is simple and accurate,
providing an accuracy not worse than sophisticated Pade and P\'ade-Borel
techniques involving about ten terms, when they are available; (iii) It is
regular, unambiguously prescribing the way of action. The method suggested
is a variant of the method of self-similar renormalization [4-6]. The latter
is a renormalization-group approach using self-similarity of subsequent
perturbative terms. Since renormalization group is nothing but a kind of a
dynamical system, the approach can be formulated in the language of
dynamical theory with the usage of its powerful techniques as well as of
those of control theory [6-8]. Then the number of the approximation order
plays the role of discrete time. Motion with respect to the latter
corresponds to transfer from one approximation to another. This makes it
possible to define a dynamical system whose trajectory is bijective to the
sequence of approximations. Such a dynamical system with discrete time has
been called the approximation cascade, Convergence of a sequence of
approximations is equivalent to stability of a dynamical trajectory. The
stability and, respectively, convergence are governed by control functions.
The fixed point of a trajectory defines the sought function.

This method has been successfully applied to the eigenvalue problem in
quantum mechanics [9,10]. Here we consider the resummation problem for the
asymptotic series in the theory of critical phenomena and advance a novel
variant of the method, more appropriate for this problem.

\section{Method of Self-Similar Renormalization}

The complete description of the method with the corresponding mathematical
foundation can be found in Refs. [4-8]. In this Section we give only the
general scheme of the method, which is necessary for explaining the new
variant we suggest.

Suppose we are interested in a function $f(x)\ $of the variable $x\in
(-\infty ,\infty ).\ $Let this function satisfy a complicated equation that
cannot be solved exactly. Assume that by means of perturbation theory we can
get a sequence $\{p_k(x)\}\ $of perturbative approximations $p_k(x),$where $%
k=0,1,2...$,\ enumerates the approximation order. Usually, perturbation
sequences are divergent. To extract a meaningful result from a divergent
sequence one has to involve the so-called resummation techniques. In the
method of self-similar renormalization a divergent sequence can be made
convergent by introducing additional functions governing convergence (see
[4-8]). These functions, because of their role are called governing or
control functions. Let $s${\it \ }be a set of such control functions
entering into a sequence $\{F_k(x,s)\}$ obtained by a perturbation algorithm.

In addition to introducing the control functions, the main idea of the
method of self-similar renormalization is to treat the passage from one
approximation to another as a motion with respect to the approximation
number $k=0,1,2$,... This motion is realized in the functional space of the
considered function as follows. Define the initial approximation

\begin{equation}
\label{1}F_0(x,s)=f 
\end{equation}
as an equation for the expansion function $x=x(f,s).\ $Substitute the latter
back to $F_k,\ $so that 
\begin{equation}
\label{2}y_k(f,s)\equiv F_k(x(f,s),s). 
\end{equation}
The relation inverse to (2) is 
\begin{equation}
\label{3}F_k(x,s)=y_k(F_0(x,s),s). 
\end{equation}

Let $\{y_k\}\ $form a group of transformations with respect to $k=0,1,2$...
Then the trajectory $\{y_k(f,s)\}\ $of this dynamical system, according to
definitions (2) and (3), is bijective, that is, in one-to-one
correspondence, to the approximation sequence $\{F_k(x,s)\}.\ $This
dynamical system with discrete time $k$ has been called [7,8] the
approximation cascade. The attracting fixed point of the cascade trajectory
is, by construction, bijective to the limit of the approximation sequence $%
\{F_k(x,s)\},\ $that is, corresponds to the sought function.

To deal with continuous time is easier than with discrete. Therefore, we
embed the approximation cascade $\{y_k\}\ $into an approximation flow $%
\{y(t,...)\}$ with continuous time $t\geq 0.${\sc \ }This implies that the
trajectory $\{y(t,f,s)\}$ of the flow passes when $\ t=k=0,1,2,...,$ through
all the points of the cascade trajectory, 
\begin{equation}
\label{4}y(k,f,s)=y_k(f,s)\ \qquad (k=0,1,2,...). 
\end{equation}
The evolution equation for the flow reads 
\begin{equation}
\label{5}\frac \partial {\partial t}y(t,f,s)=\text{{\it v}}(y(t,f,s)), 
\end{equation}
with the right-hand side being the velocity field. The latter, in the
language of renormalization-group theory, is often called the Gell-Mann-Low
or $\beta -$function.

Integrating the evolution equation (5) from $\ t=k$ to $t=k^{*}$, we get the 
{\it evolution integral}

\begin{equation}
\label{6}\int_{y_k}^{y_{k+1}^{*}}\frac{df}{\text{{\it v}}(f,s)}=k^{*}-k, 
\end{equation}
in which $\ y_k=y(k,f,s)$ and\ \ $y_{k+1}^{*}=y(k^{*},f,s).$ Before
specifying the numbers $k$ and $k^{*}$ in the limits of the evolution
integral, let us note that the differential form (5) of the evolution
equation, or its integral form (6), are equivalent to the functional
relation 
\begin{equation}
\label{7}y(t+t^{\prime },f,s)=y(t,y(t^{\prime },f,s),s). 
\end{equation}
The latter in physical applications is labeled as the self-similarity
relation, which explains the term we use. The self-similarity, in general,
can occur with respect to motion over different parameters. In our case,
this is the motion over the steps of a calculational procedure, the number
of steps playing the role of effective time.

If there exists an attractive fixed point of the approximation-flow
trajectory, then it is always possible to find a number $k^{*}$ in the
evolution integral (6) such that the upper limit $y_k^{*}\ $ would
correspond to an expression 
\begin{equation}
\label{8}F_k^{*}(x,s)\equiv y(k^{*},F_0(x,s),s) 
\end{equation}
representing, with the desired accuracy, the sought function\ $f(x).\ $If $\
y_k^{*}$ would be an exact fixed point, then (8) would give an exact answer
to the problem. However, a fixed point can be reached only after infinite
number of steps $k\rightarrow \infty .$ For a finite number $k$, the limit $%
y_k^{*}$ may represent a fixed point approximately, because of which it is
named the quasi -fixed point. Our aim is to reach the latter as fast as
possible, that is, during the minimal time 
\begin{equation}
\label{9}t_k^{*}=\text{min}(k^{*}-k), 
\end{equation}
or the minimal number of steps. When there are no additional restrictions,
the minimal number of steps counted by $k$ is $1,$ so that 
\begin{equation}
\label{10}abs\ \text{min}\ t_k^{*}=1. 
\end{equation}
In the case when some constraints are imposed on the motion, the minimal
time (9) should correspond to the conditional minimum. For instance, if a
value $f_0\equiv f(x_0)$ of the sought function $f(x)$ is given for some $%
x_{0,}$ then we can find $t_k^{*}$ by requiring the trajectory of the
approximation cascade to pass through the given point $f_{0.\text{ }}$.

To calculate the evolution integral (6), we need to define the velocity
field. This can be done by the Euler discretization of (5) yielding the
finite-difference form 
\begin{equation}
\label{11}\text{{\it v}}_k(f,s)=y_k(f,s)-y_{k-1}(f,s). 
\end{equation}
Substituting (11) into (6), and using (3), we come to the representation 
\begin{equation}
\label{12}\int_{F_k}^{F_{k+1}^{*}}\frac{df}{\text{{\it v}}_{k+1}(f,s)}%
=t_k^{*}, 
\end{equation}
for the evolution integral (6), where $F_k=F_k(x,s),\
F_{k+1}^{*}=F_{k+1}^{*}(x,s).$

Finally, we have to define the set $s$ of control functions. The role of the
latter is to govern the convergence of the approximation sequence. This
convergence can be expressed, in the language of dynamical theory, as the
stability of the cascade trajectory. A useful tool for analyzing stability
is a set $\{\mu _k\}$ of the local multipliers 
\begin{equation}
\label{13}\mu _k(f,s)=\frac \partial {\partial f}y_k(f,s). 
\end{equation}
The inequality 
\begin{equation}
\label{14}\mid \mu _k(f,s)\mid <1 
\end{equation}
is the condition of local stability at the step $k$ with respect to the
variation of an initial point\ $f.$ The equality $\mid \mu _k(f,s)\mid =1$
implies local neutral stability. For a convergent sequence corresponding to
a contracting mapping, the condition of asymptotic stability is 
\begin{equation}
\label{15}\mid \mu _k(f,s)\mid \ \rightarrow 0\qquad (k\rightarrow \infty ). 
\end{equation}
\qquad The approximation cascade $\{y_k\}$ describes the motion in the
functional space $\{f\}.\ $To return to the domain $\{x\},\ $we must use the
inverse transformation (3). With the help of the latter we may pass from the
multiplier (13) given on the functional space $\{f\}$ to its image 
\begin{equation}
\label{16}m_k(x,s)=\mu _k(F_0(x,s),s) 
\end{equation}
being a function of $x.\ $For the image (16), the same stability condition
as in (14) can be written, 
\begin{equation}
\label{17}\mid m_k(x,s)\mid <1. 
\end{equation}
According to (15), the local multipliers diminish when approaching an
attracting fixed point. That is, the variation of initial condition \ $f$
produces lesser and lesser effect on the trajectory as soon as the attractor
becomes closer. In other words, the lesser are the absolute values of
multipliers, the more stable is the trajectory. Therefore, it is reasonable
to define the control functions as those minimizing the absolute values of
the local multipliers, making by this the trajectory more stable at each
step $k.$ In this way, a set $s$ of control functions is to be defined by
the {\it principle of maximal stability}{\bf \ }written as 
\begin{equation}
\label{18}\mid m_k(x,s_k(x))\mid ={\text{min}}_s\mid m_k(x,s)\mid .\text{ } 
\end{equation}
Because of this, the control functions $s_k(x)$ defined by the principle
(18) may be called the {\it stabilizing}{\bf \ {\it functions} }or {\it %
stabilizers.}

Note that the control functions may be introduced in several ways, as is
discussed in Refs.[4-10], however always being related to stability
conditions and the closeness of a trajectory to an attracting fixed point.
In all the cases the control functions are to be defined so that they could
accomplish their main job, i.e., to govern the convergence of an
approximation sequence, which, in the terms of dynamical theory is
equivalent to stabilizing the cascade trajectory. In the present paper we
shall use the definition of stabilizers given in (18).

After the stabilizers are defined, we have to substitute them into the
corresponding approximations $F_k(x,s)$ getting 
\begin{equation}
\label{19}f_k(x)\equiv F_k(x,s_k(x)). 
\end{equation}
This stage can be called the {\it stabilizing renormalization }of a
perturbative sequence.

Then, considering the motion near the renormalized quantity (19) by means of
the evolution integral (12), we obtain 
\begin{equation}
\label{20}f_k^{*}(x)\equiv F_k^{*}(x,s_k(x)). 
\end{equation}
This step can be called the{\it \ dynamical renormalization}{\bf .\ }And the
whole procedure of the double renormalization (19) and (20) is named the
self-similar renormalization. It is worth noting that the evolution equation
(5) is, generally, nonlinear and can have several different solutions
leading to different self-similar approximations (20). In such a case, to
select a physically meaningful solution, we need to involve additional
conditions as constraints.The role of the latter can be played, e.g., by
properties of symmetry, asymptotic properties at $x\rightarrow 0$ or $%
x\rightarrow \infty ,\ $sum rules or other relations containing some known
information on the character of the sought solution. Such additional
constraints narrow the set of possible solutions to a class with desired
properties. Thus, we should always remember from what class we are looking
for a solution.

Keeping in mind that we wish to get a good accuracy for the sought function,
having just a few perturbative terms available, we need to find out some
tricks which could effectively increase perturbation order. We suggest below
one such trick.

Suppose that there is a sequence of approximations $p_k(x)$ having
polynomial structure, $k$ showing the order of the polynomial. This order
can be effectively increased by means of the multiplicative transformation 
\begin{equation}
\label{21}P_k(x,s)=x^s\ p_k(x),\ s\geq 0. 
\end{equation}
Then, the order of the expression (21) becomes $k+s.$ The transformation
inverse to (21), as is evident, is 
\begin{equation}
\label{22}p_k(x)=x^{-s}\ P_k(x,s). 
\end{equation}
Following the method described above, we consider the sequence $\{P_k(x,s)\}$
and construct an approximation cascade $\{y_k\}$ whose trajectory $%
\{y_k(f,s)\}$ is bijective to $\{P_k(x,s)\}.$ Solving the evolution integral
(12), we have $P_k^{*}(x,s).$ From the principle of maximal stability (18)
we define the stabilizers $s_k(x).$ Substituting these into $P_k^{*}(x,s)$
and invoking the inverse transformation (22), we obtain the self-similar
approximation 
\begin{equation}
\label{23}f_k^{*}(x)=x^{-s_k(x)}\ P_k^{*}(x,s_k(x)). 
\end{equation}
The multiplicative transformation (21) looks as the most natural for the
case when the perturbative approximations $p_k(x)$ have the form of
polynomials or series, generally speaking, not necessarily in integer
powers. The factor $x^s$ effectively increases the approximation order, and $%
s$ plays simultaneously the role of stabilizer.

What power $s$ we have to choose, that is, to what effective order we need
to go is dictated by the principle of maximal stability selecting the most
stable trajectory of the approximations cascade. In particular, it may
happen that $s=0$, and we do not need to proceed further, or, vice versa, we
may have to go to the limit of $s\rightarrow \infty $, thus making allowance
for all approximation orders. In each concrete case, an effective order
which we need to reach depends on how good is the perturbative sequence $%
\{p_k(x)\}$ we start with and, respectively, how much information can be
extracted from its first terms by means of the double renormalization (19)
and (20).

The optimization by introducing the stabilizing control functions into the
powers of perturbative polynomials renormalizes the algebraic structure of
the latter. Because of this, and in order to distinguish the suggested
optimization procedure from other possible variants, we shall call it the 
{\it algebraic}{\bf \ {\it self-similar renormalization}.}

To concretize the procedure, let us write explicitly 
\begin{equation}
\label{24}p_k(x)=\sum_{n=0}^ka_n\ x^n,\ \ a_n\neq 0, 
\end{equation}
as a polynomial of the order $k.$ Following (21) define 
\begin{equation}
\label{25}P_k(x,s)=\sum_{n=0}^ka_n\ x^{n+s}. 
\end{equation}
Similarly to (1), we have 
\begin{equation}
\label{26}P_0(x,s)=a_0\ x^s=f, 
\end{equation}
from where the expansion function is 
\begin{equation}
\label{27}x(f,s)=(\frac f{a_0})^{1/s}. 
\end{equation}
The definition (2) yields the points 
\begin{equation}
\label{28}y_k(f,s)=\sum_{n=0}^ka_n\ (\frac f{a_0})^{n/s+1}\ 
\end{equation}
of the approximation-cascade trajectory. For the velocity field (11) we get 
\begin{equation}
\label{29}\text{{\it v}}_{k+1}(f,s)=a_{k+1}(\frac f{a_0})^{\frac{k+1}s+1}. 
\end{equation}
From the evolution integral (12) we find 
\begin{equation}
\label{30}P_{k+1}^{*}=\frac{P_k}{(1-\frac{(k+1)\ a_{k+1}\ t_k^{*}}{s\ a_0^{
\frac{k+1}s+1}}P_k^{\frac{k+1}s})^{\frac s{k+1}}}. 
\end{equation}
The multiplier (13) becomes 
\begin{equation}
\label{31}\mu _k(f,s)=\sum_{n=0}^k\frac{a_n}{a_0}\ (1+\frac ns)\ (\frac
f{a_0})^{\frac ns}, 
\end{equation}
and its image (16) reads 
\begin{equation}
\label{32}m_k(x,s)=\sum_{n=0}^k\frac{a_n}{a_0}\ (1+\frac ns)\ x^n. 
\end{equation}
The principle of maximal stability (18) defines the stabilizers $s_k(x),\ $%
whose explicit expressions depend on the coefficients $a_n.\ $According to
the transformations (21)-(23), from (30) we obtain 
\begin{equation}
\label{33}f_{k+1}^{*}=\frac{p_k(x)}{(1-\frac{(k+1)\ a_{k+1}\ t_k^{*}}{s\
a_0^{\frac{k+1}s+1}}x^{k+1}p_k(x)^{\frac{k+1}s})^{\frac s{k+1}}}, 
\end{equation}
where $s_k(x)$ defines the most stable trajectory. When there are no
additional conditions, the minimal value $t_k^{*}=1,$ as in (10).

As is noted above, it may happen that the most stable trajectory corresponds
to $s\rightarrow \infty .$ Let us show how the self-similar approximation
(33) simplifies in this case. It is straightforward to check that the limit
of the right side in (33), as $s\rightarrow \infty ,\ $leads to 
\begin{equation}
\label{34}f_{k+1}^{*}(x)=p_k(x)\exp (\frac{a_{k+1}}{a_0}x^{k+1}). 
\end{equation}
One may notice that renormalizing $p_k(x)$ in (34) we can obtain the
recurrence relation 
\begin{equation}
\label{35}f_{k+1}^{*}(x)=f_k^{*}(x)\exp (\frac{a_{k+1}}{a_0}x^{k+1}). 
\end{equation}
It is possible also to derive several other relations permitting to repeat
the self-similar renormalization several times, which is useful when
working with high-order terms. However, in what follows we shall limit the
consideration of particular examples by keeping only a few terms of the
corresponding perturbative series.\ This is to emphasize that the method
suggested allows to reach good accuracy with a minimal number of
perturbative terms, when no other resummation technique is applicable.
Comparing (33) with (34), we see that the self-similar renormalization can
yield quite different expressions, from the fractional form to exponential
one.\ Below we shall illustrate this by some simple examples. Each appearing
form of an approximation results from choosing the most stable trajectory by
which it is possible to reach a quasi-fixed point during the minimal time.
Recall in this connection the analogy with classical mechanics. Notice also
that it is possible to follow a trajectory which is stabilized by imposing
additional conditions, such as asymptotic properties, or prescribing that
the trajectory is to pass through some given points. In such a case the
motion will not, generally, be accomplished during the absolute minimal time
(10), but the latter should be defined from the additional constraints
imposed. All these variants will be exemplified in the following sections.

\section{Illustration by Simple Examples}

Suppose that by perturbation theory we have got 
\begin{equation}
\label{36}p_1(x)=1-x, 
\end{equation}
with $0\leq x\ll 1.$ How could one continue this expression from small $x\ll
1$ to $x\geq 1,\ $when no other information is available?

Following the algebraic self-similar renormalization, we construct the
transformed polynomial (21) or (25) , which for the case (36) is 
\begin{equation}
\label{37}P_1(x,s)=x^s-x^{1+s}. 
\end{equation}
According to (26), we have the expansion function 
\begin{equation}
\label{38}x(f,s)=f^{1/s}. 
\end{equation}
Then, Eq.(28) gives 
\begin{equation}
\label{39}y_1(f,s)=f-f^{1+\frac 1s}. 
\end{equation}
The velocity field (29) becomes 
\begin{equation}
\label{40}\text{{\it v}}_1(f,s)=-f^{1+\frac 1s}. 
\end{equation}
The evolution integral (12), leading to (30) now yields 
\begin{equation}
\label{41}P_1^{*}(x,s)=(\frac{sx\ }{s+x})^s. 
\end{equation}
For the multiplier (32) we have 
\begin{equation}
\label{42}m_1(x,s)=1-(1+\frac 1s)\ x. 
\end{equation}
Minimizing the absolute value of the latter gives the stabilizer 
\begin{equation}
\label{43} 
\begin{array}{c}
s_1(x)=\frac x{1-x},\qquad 0\leq x\leq 1, \\ 
s_1(x)\rightarrow \infty ,\qquad x\geq 1. 
\end{array}
\end{equation}
The self-similar approximation (33) reduces to 
\begin{equation}
\label{44}f_1^{*}(x)=(\frac{s_1(x)}{s_1(x)+x})^{s_1(x)}. 
\end{equation}
Being interested in the region $x\geq 1$ we have to take the limit $%
s_1(x)\rightarrow \infty .$ Therefore, for the self-similar approximation
(44) we obtain 
\begin{equation}
\label{45}f_1^{*}(x)=\lim _{s\rightarrow \infty }(\frac{s\ }{s+x})^s=e^{-x}. 
\end{equation}
In the same way, a linear expansion 
\begin{equation}
\label{46}p_1(x)=a_0+a_1x, 
\end{equation}
with \ $a_{0,}\ a_1$ $\neq 0,\ $derived for $\mid x\mid \ll 1,$ can be
continued to the region $\mid x\mid \geq \mid \frac{a_0}{a_1}\mid $ where it
is represented by the self-similar approximation 
\begin{equation}
\label{47}f_1^{*}(x)=a_0\exp (\frac{a_1}{a_0}x). 
\end{equation}
Thus, we may conclude that the exponential (47) is a general self-similar
representation of a linear expansion (46), when {\it no additional
constraints are imposed.}

Now turn to the case, when we want to construct a self-similar continuation
of (36) satisfying the prescribed asymptotic behavior 
\begin{equation}
\label{48}f(x)\propto x^n,\qquad x\rightarrow \infty , 
\end{equation}
where $n>0$ is fixed. Repeating the same procedure as earlier, we come to
(44). Comparing the latter with (48), we get \ $s_1(x)=n,$ so that 
\begin{equation}
\label{49}f_1^{*}(x)=(\frac{n\ }{n+x})^n. 
\end{equation}
Generalizing this result for a linear combination (46) under the asymptotic
condition (48), we have 
\begin{equation}
\label{50}f_1^{*}(x)=a_0(\frac{na_0\ }{na_{0\ }-a_1x})^n. 
\end{equation}
In this way, one perturbative expansion may have several self-similar
representations corresponding to different imposed constraints. The form of
these representations can vary between the exponential, (47), and fractional
one, (50). However, for each given constraint this form is uniquely defined.
If no constraints are imposed, the form of the resulting self-similar
approximation is governed by the stabilizers obtained from the principle of
maximal stability of a self-similar trajectory.

Let us illustrate how accurate is a self-similar approximation and how it is
possible to increase the accuracy by considering higher-order terms of a
perturbative expansion. For this purpose take the function $\ \ln (1+x)$
with $\ x\geq 0$. We opt for this function, as an example, since the
logarithmic expressions are typical of thermodynamic potentials in
statistical mechanics and of generating functionals in field theory.

Write down the three first perturbative approximations for $\ln (1+x)$ in
powers of $x,\ $thinking that $x\ $is small: 
\begin{equation}
\label{51} 
\begin{array}{c}
p_1(x)=x, \\ 
p_2(x)=x- 
\frac{x^2}2, \\ p_3(x)=x-\frac{x^2}2+\frac{x^3}3. 
\end{array}
\end{equation}
Our aim is to construct self-similar approximations for $\ln (1+x)$ in the
region $x\approx 1,$ with expansions (51) that are valid only for $x\ll 1.$

Following the standard prescription of the method, define the transformed
polynomials (25) for those in (51), which gives 
\begin{equation}
\label{52} 
\begin{array}{c}
P_1(x,s)=x^{1+s}, \\ 
P_2(x,s)=x^{1+s}- 
\frac{x^{2+s}}2, \\ P_3(x,s)=x^{1+s}-\frac{x^{2+s}}2+\frac{x^{3+s}}3. 
\end{array}
\end{equation}
Now, the initial, i.e, the lowest order approximation is $P_1(x,s).$
Therefore, as in (26), from the equation $P_1(x,s)=f$ \ we find the
expansion function 
\begin{equation}
\label{53}x(f,s)=f^{\frac 1{1+s}}. 
\end{equation}
The points of the approximation -cascade trajectory (28) are 
\begin{equation}
\label{54} 
\begin{array}{c}
y_1(f,s)=f, \\ 
y_2(f,s)=y_1(f,s)-\frac 12\ f^{
\frac{2+s}{1+s}}, \\ y_3(f,s)=y_2(f,s)+\frac 13\ f^{\frac{3+s}{1+s}}. 
\end{array}
\end{equation}
For the velocity field (21) we get 
\begin{equation}
\label{55} 
\begin{array}{c}
\text{{\it v}}_2(f,s)=-\frac 12\ f^{\frac{2+s}{1+s}}, \\ \text{{\it v}}%
_3(f,s)=\frac 13\ f^{\frac{3+s}{1+s}}. 
\end{array}
\end{equation}
The evolution-integral solutions in (30) become 
\begin{equation}
\label{56}P_2^{*}(x,s)=[\frac{2\ (1+s)\ x}{2\ (1+s)+x}]^{1+s},\quad
P_3^{*}(x,s)=[\frac 1{x^{2\ }(1-\frac x2)^{\frac 2{1+s}}}-\frac 2{3(1+s)}]^{
\frac{1+s}2}. 
\end{equation}
And for the multipliers in (32) we have 
\begin{equation}
\label{57} 
\begin{array}{c}
m_2(x,s)=1-\frac 12( 
\frac{2+s}{1+s})\ x, \\ m_3(x,s)=m_2(x,s)+\frac 13(\frac{3+s}{1+s})\ x^2. 
\end{array}
\end{equation}
The stabilizers are to be defined at each step by minimizing the absolute
values of the corresponding multipliers in (57). For instance
$$
\begin{array}{c}
s_2(x)=0,\qquad 0\leq x\leq 1, \\ 
s_2(x)=2 
\frac{x-1}{2-x},\qquad 1\leq x\leq 2, \\ s_2(x)=\infty ,\qquad x\geq 2. 
\end{array}
$$

The corresponding expression for $s_3(x)$ is also easy to find. For the
region of interest, where $x\approx 1,\ $ we have
$$
s_2(x)=0,\qquad s_3(x)\rightarrow \infty \qquad (x=1). 
$$

This leads to the self-similar approximations 
\begin{equation}
\label{58} 
\begin{array}{c}
f_2^{*}(x)= 
\frac{2x}{1+x}, \\ f_3^{*}(x)=x(1-\frac x2)\exp (\frac{x^2}2), 
\end{array}
\end{equation}
obtained from (56) as in (23).

In order to check the accuracy of (58) as compared to the perturbative
expansions (51), define the percentage errors 
\begin{equation}
\label{59}\epsilon _k(x)\equiv \frac{p_k(x)-f(x)}{\mid f(x)\mid }\times
100\% 
\end{equation}
and , respectively, 
\begin{equation}
\label{60}\epsilon _k^{*}(x)\equiv \frac{f_k^{*}(x)-f(x)}{\mid f(x)\mid }%
\times 100\%, 
\end{equation}
where $f(x)=\ln (1+x).$ At the point $x=1,\ $the errors (59) and (60),
calculated with respect to $f(1)=\ln (2)=0.693,$ are%
$$
\begin{array}{c}
\epsilon _2(1)=-28\%,\quad \epsilon _3(1)=20\%, \\ 
\epsilon _2^{*}(1)=-3.8\%,\quad \epsilon _3^{*}(1)=0.67\%. 
\end{array}
. 
$$

As is seen, the accuracy of the self-similar approximations in (58) is an
order higher than that of the perturbative expansions in (51), and this
accuracy can be increased by taking into account additional perturbative
terms.

In this section we considered simple examples in order to make transparent
all steps of our method. This will permit us in the following sections to
avoid the repetition of the technical details when applying the method to
more complicated physical problems.

\section{Calculation of Critical Temperature for 2d and 3d Ising Model}

In this Section we calculate the critical temperature $T_c\ $of the
two-dimensional $(2d)\ $and three-dimensional $\ (3d)\ $Ising model starting
from approximate expressions for$\ T_c\ $obtained by the
variational-cumulant expansion (VCE)[11]. The convergence of VCE
approximations is very slow and even using the seven consecutive
approximations one obtain $T_c$ for $2d$ Ising model with the percentage
error of about $11\%$. We use below the simple variant of the self-similar
renormalization when the exact value of the sought function is known for
some point, namely the knowledge of $T_c$ for the $2d$ Ising model will be
used as an optimization condition for the trajectory, determining the
optimal number of steps $t^{*}$, then used for calculating $T_c$ for the $3d$
Ising model. Although the expressions below are a little complicated for
getting the result in an explicit form, we can realize here the numerical
variant of the self-similar renormalization , when sought function is
obtained implicitly.

We rewrite the expressions for the critical temperature from [11] in terms
of a new variable $x=\frac 1d$ , where $d$ is the space dimension. Then we
calculate $\widetilde{T_c}(x),$ related to $T_c$ as follows: $T_c=\frac 1x$ $
\widetilde{T_c}(x).$

Write down the three approximations to $\widetilde{T_c}(x),$%
\begin{equation}
\label{61} 
\begin{array}{c}
\widetilde{T_{c1}}(x)=2-x, \\ \widetilde{T_{c2}}(x)=\frac{12-12x+2}{6-3x},
\\ \widetilde{T_{c3}}(x)=\frac{24-36x+8x^2+5x^3}{12-12x+2x^2}, 
\end{array}
\end{equation}
which correspond to the second, third and fourth approximations of [11],
respectively.

Following the standard approach described above in Section 3, from the
equation $\widetilde{T_c}(x)=f$ we find the expansion function $x(f)=2-f.$
The points of the approximation cascade trajectory are 
\begin{equation}
\label{62} 
\begin{array}{c}
y_1(f)=f, \\ 
y_2(f)=\frac 23( 
\frac{f^2+2f-2}f), \\ y_3(f)=-\frac 12(\frac{5f^3-38f^2+56f-24}{f^2+2f-2}). 
\end{array}
\end{equation}
For the velocity field we get%
$$
\text{{\it v}}_3(f)=-\frac 16\frac{(19f^2-22f+4)}{(f^2+2f-2)}\frac{(f-2)^2}%
f. 
$$

The evolution integral cannot be calculated explicitly, so we expressed $
\widetilde{T}_{c3}^*$ implicitly as a function of $x$ and $t^{*}$ given 
by the equation%
$$
\int_{\widetilde{T}_{c2}}^{\widetilde{T_{c3}}^{*}}\frac{df}{\text{{\it v}}%
_3(f)}=t^{*}, 
$$
and obtained $\widetilde{T}_{c3}^{*}$numerically for two different values $\
t^{*}=1,$ or $t^{*}=1.5515,$ corresponding to a non-optimized and optimized
variant respectively. The latter, optimizing number of steps, was obtained
from the condition $\widetilde{T}_{c3}^*(\frac 12,t^{*})=\frac{T_{2d}}2,$
where $T_{2d}=2.269$ is the celebrated Onsager temperature for the $2d$
Ising model.

At $\ t^{*}=1$ and $d=2$ we found that $T_c^{*}=2.531$ and the percentage
error of our estimate is $\ \epsilon _3^{*}(T_c)=11.542\%,\ $approaching the
percentage error reached using seven consecutive approximations to $T_c\ $%
[11].

We should point out here that the authors of [11] did not attempt to
calculate $T_c\ $ for $3d$ Ising model, where the best known ''exact''
numerical value of the critical temperature is $T_c=4.51$ [12]. We obtained $%
T_c^{*}=4.712$ \ at $\ t^{*}=1$ and $\ d=3.$ The percentage error is $%
\epsilon _3^{*}(T_c)=4.47\%.$ \ For the optimized $\ t^{*}=1.5515$ our
estimate of $\ T_c^{*}=4.548\ $ with the error equal to $\ 0.838\%$ \ is
quite accurate.

It is worth noting that, in principle, the parameter $x=\frac 1d$ \ may be
considered as ''small'' for $d=3$ and the expressions for $\widetilde{T_c}%
(x) $can be expanded in powers of $x,$ thus presenting $T_c$ in the form of
''$\frac 1d-$expansion''. We performed the same renormalization procedure as
above for%
$$
\begin{array}{c}
\widetilde{T_{c2}}(x)\approx 2-x-\frac{x^2}6, \\ \widetilde{T_{c3}}%
(x)\approx 2-x-\frac{2x^2}3-\frac{x^3}{12}, 
\end{array}
$$
and found that in $2d$ \ for $t^{*}=1,\;\; T_c^{*}=2.453$ \ with\ $\epsilon
_3^{*}(T_c)=8.109\%$ and in $3d$-case with \ $t^{*}=1,\ $we have $%
T_c^{*}=4.701$ \ with \ $\epsilon _3^{*}(T_c)=8.109\%.$\ 

For the optimized $\ t^{*}=1.241$ and $\ d=3,\; \; T_c^{*}=4.624$ \ and \ $%
\epsilon _3^{*}(T_c)=2.527\%$. We see that the renormalized $\ \frac 1d-$%
expansion gives by order of magnitude the same accuracy that is reached from
the renormalized original expressions.

One can conclude from the results presented in this Section, that both the
rate of convergence and accuracy of the VCE are greatly improved by applying
the self-similar renormalization to the starting VCE approximations.
Situation encountered while renormalizing $\ T_c$ was somewhat very simple,
since we possessed three reasonable approximations for renormalization and
also knew the exact value of $T_c$ \ for the $2d$ Ising model. In the next
Section we meet the case when the number of the terms available are not
sufficient for any variant of renormalization discussed above and no exact
value for the quantity under consideration is known.

\section{Renormalization of Shifted-Power Expansions}

In this Section we apply a modified variant of the self-similar
renormalization to the ''shifted-power expansions'' for the critical indices
[13]. Shifted-power expansion or $\Delta -$expansion is of particular
interest for systems without upper critical dimensionality. It can be
applied to the calculation of critical exponents of a system described by a
Landau-Ginzburg (LG) Hamiltonian. Within the framework of one of many
possible realizations of the method, the leading nonlinear term of the LG
Hamiltonian,$\psi ^4,$ is replaced by $(\psi ^2)^{3-\Delta },$ then $\Delta $
is used as an expansion parameter and at the end of calculations one should
set $\Delta =1.$ The expressions for critical indices $\eta $ and $\nu $
were obtained in the following form: 
\begin{equation}
\label{63}
\begin{array}{c}
\eta =b_2(n)\ \Delta ^2,\qquad b_2(n)=
\frac{(n+4)(n+2)}{48(3n+22)^2}, \\ \nu =a_0+a_2(n)\ \Delta ^2,\qquad
a_0=\frac 12,\ a_2(n)=\frac{(n+4)(n+2)}{12(3n+22)^2},
\end{array}
\end{equation}
where $n$ is the number of the components of the order parameter.

The corrections to the mean-field values given by (63) are about two orders
of magnitude too small. The authors of [13] had noticed that the results are
strongly influenced by the $\psi ^6$ interaction and remain too distant from
the analyzed $\psi ^4\ $behavior. This conclusion agrees well with the
rigorous results of [14] where it was shown that special Gaussian points \ $%
n=-4,-2$ \ should appear when the $\psi ^6$ model is considered, while for
the $\psi ^4$ model only the point $n=-2$ \ exists. Appearance in the
expressions (63) of the combination \ $n+4$ signalizes that the critical
indices in the interesting physical region $n=0,1...3$ are influenced by the
point that does not have any meaning for the $\psi ^4\ $model at all.
Nevertheless, the Gaussian point $n=-2$ does have physical meaning for the $%
\psi ^4$ model [15-17] and, formally, the results given by (63) are correct
at $n=-2.$ Therefore, one may hope that a systematic and pernicious
influence of the point $n=-4$ can be weakened by some renormalization
procedure, at least when a physical region not very distant from the point $%
n=-2$ is considered. We should stress also that the expressions in (63) do
not obey another rigorously studied limit of the LG Hamiltonian, when $%
n\rightarrow \infty ,\ $i.e. $\eta (n\rightarrow \infty )\neq 0$ and $\ \nu
(n\rightarrow \infty )\neq 1$ [18]. Therefore we do not expect that a
successful renormalization of (63) could be realized for $n$ very distant
from the point $n=-2.$

Our approach to the renormalization of $\eta $ and $\nu $ should vary, since
for the index $\nu $ the two terms in the $\Delta -$expansion are available
and the renormalization could be carried on straightforwardly, while for the
index $\eta $ only single term was obtained and it is not possible to
proceed with extra assumptions.

Write down the two consecutive approximations to the index $\nu $ in powers
of $\Delta :$%
$$
\begin{array}{c}
\nu _0(n)=a_0, \\ 
\nu _2(n)=a_0+a_2(n)\ \Delta ^2, 
\end{array}
$$
and apply the same procedure that leads to the expression (47) with a
substitution of $x$ to $\Delta ^2.$ We also retain in the final expression
for the renormalized critical index $\nu _2^{*}$ an effective time $t^{*}$
which will be exploited as an optimization parameter: 
\begin{equation}
\label{64}\nu _2^{*}(\Delta ,n,t^{*})=a_0\exp \{\frac{a_2(n)}{a_0}\Delta
^2t^{*}\}. 
\end{equation}
Setting here $\Delta =1$ and $\ t^{*}=1$ we see that no considerable effect
was achieved and the index $\nu $ remains in the interval $\ \nu =0.5$ to $%
0.509$ \ while $n$ varies from $n=-2$ \ to infinity. Impose now an
additional condition that in the case of $n=0,$ corresponding to the random
walk problem [19], the approximation cascade trajectory pass through the
value of the critical index $\nu =0.588\ $ known approximately, but with
very high accuracy for this physical problem [20]. The choice of this point
for optimization of the trajectory is dictated also by the desire to receive
renormalized values for the physically interesting region $n=1,2,3$ \ of
''true'' phase transitions using information only from the physically
distant region, where the random walk problem may be a good choice because
it does not correspond to a ''true'' phase transition, but only a formal
analogy exists with the $n\rightarrow 0$ limit of the LG Hamiltonian.

So, from the condition $\nu _2^{*}(1,0,t^{*})=0.588$ \ we obtain \ $%
t^{*}=59\ $ and the following values at $n=1,2;3:$%
$$
\nu _2^{*}(1,1,t^{*})=0.633,\quad \nu _2^{*}(1,2,t^{*})=0.676,\quad \nu
_2^{*}(1,3,t^{*})=0.715. 
$$

These values are reasonable as compared to the experiment, high-temperature
series, and Borel-summation results [ 21]. We should point out that only by
means of a single parameter we obtain simultaneously renormalized values for
the physically interesting situation, i.e a systematic multiplicative error
in the initial expansion, caused by peculiarities of the $\Delta -$expansion
can be eliminated by a single renormalization step.

It is also possible to find optimal \ $t^{*}$ from the condition restoring
the correct value of $\nu $ at $n\rightarrow \infty ,$ where, as is shown
rigorously, $\nu =1.$ In our case, the results happen to be much better than
for the initial $\Delta -$expansion, but still largely underestimate $\nu ,\ 
$e.g at $n=0,\ \nu ^{*}=0.544$. Nevertheless , this variant of optimization
is of general interest because often the limit $\ n\rightarrow \infty \ ($or 
$d\rightarrow \infty )$ is well known.

In the case of the index $\eta $ one should proceed differently, since only
a single term in the$\ \Delta -$expansion is known: we added to the
expression for $\eta $ the term linear in $\Delta $ with some yet unknown
positive coefficient $b$ defining thus a new quantity $\widetilde{\eta .}\ \ 
$Then carried out the renormalization procedure for $\widetilde{\eta },$
repeating the steps leading to Eq.35. From the renormalized quantity $
\widetilde{\eta }^{*}$ using the variational condition $\frac{\partial 
\widetilde{\eta }^{*}}{\partial b}=0,$ we determine $b$ as a function of $%
n,\ \Delta ,\ t^{*}.$ Finally, we subtracted the term $\ b\Delta \ $ from $
\widetilde{\eta }^{*}$ to find $\eta ^{*}.$ Following this prescription
define%
$$
\widetilde{\eta }=b\Delta +b_2(n)\Delta ^2, 
$$

then find 
$$
\widetilde{\eta }^{*}=b\Delta \exp \{\frac{b_2(n)}b\Delta \ t^{*}\}. 
$$

This yields 
$$
\eta ^{*}=\widetilde{\eta }^{*}-b\Delta =t^{*}b_2(n)(e-1)\ \Delta ^2. 
$$

At \ $t^{*}=1$ the results still remain too small. Imposing an additional
optimizing condition on $t^{*}$ by analogy with the case of $\nu ,\ $that $%
\eta ^{*}$ should equal to $0.026,$ where the critical index $\eta $ for the
random walk problem is taken from [20], we find that \ $t^{*}=44\ $ and 
$$
\eta ^{*}(n=1,t^{*})=0.038,\quad \eta ^{*}(n=2,t^{*})=0.048,\quad \eta
^{*}(n=3,t^{*})=0.057. 
$$

These results overestimate $\eta ,$ especially for $n=3,$ but are much more
realistic than the initial value $\eta \sim 10^{-4}-10^{-3}.$ This
systematic error can be understood if to notice that already the initial $\
\Delta -$expansion does not obey the limit of \ $n\rightarrow \infty ,$ and
\ $\eta (\infty )$ $=0.002$ \ instead of zero. This systematic deviation
cannot be fully corrected by a variational-renormalization procedure. The
same is true in the case of index $\nu ,$ but in this case more information
is available from the initial $\ \Delta -$expansion and the results of
renormalization remain reasonable even at $n=3.$

We conclude that by applying the self-similar renormalization to the $\Delta
-$expansion for the critical indices, one can obtain reasonable estimates
for $\eta $ and $\nu ,$ although further improvement of these estimates does
not seem plausible, since initial expressions violate an exact relation in
the $n\rightarrow \infty $ limit and possess an unphysical Gaussian point at 
$\ n=-4.$

In the next Section we meet the case when both limits at $\ n=-2$ and $\
n\rightarrow \infty $ are violated.

\section{Importance of Asymptotic Properties}

We have seen in the previous Section that the $\Delta -$expansion mimicking
the widely accepted Wilson $\epsilon -$expansion is, in the best case, a
crude approximation for the critical indices, since an important $%
n\rightarrow \infty $ limit is violated already in the starting terms of the 
$\Delta -$expansion. The question naturally arises whether the $\epsilon -$%
expansion obeys exact limits for critical indices, namely at $\ n=-2$ and $\
n\rightarrow \infty $ ? The discussion of this question for all critical
indices will be presented later. In this Section, we consider the Wilson $%
\epsilon -$expansion for the critical index $\delta ,$ discuss its $n=-2$
and $n\rightarrow \infty $ limits, observe that they are violated and
suggest the self-similar renormalization approach allowing to restore the
correct limiting values for the index $\delta .$

Consider the well-known Wilson $\epsilon -$expansion for the critical index $%
\delta $ [1,18] up to the quadratic terms in $\epsilon :$%
\begin{equation}
\label{65}\delta =3+\epsilon +c_2(n)\epsilon ^2,\qquad c_2(n)=\frac 12\frac{%
n^2+14n+60}{(n+8)^2}. 
\end{equation}
At $n=-2$ $\ $and $\ n\rightarrow \infty ,\ $ $\delta =4.5.$ From the exact
results for $n=-2$ vector model [16] \ and from the scaling law for the $d=3$
the result $\delta =5$ \ follows. The same value $\delta =5$ was obtained in
the case of the spherical model\ $(n\rightarrow \infty ,\ d=3)$ [18]. The
percentage error for the critical index $\delta $ in these limits is
therefore $-10\%.$ In the physical region $n=0,1...3$ \ the $\epsilon -$%
expansion gives $\delta =4.47-4.46.$ Unfortunately, there is no much
experimental data or theoretical estimates available for the index $\delta ,$
but if we accept the scaling laws as correct and estimate from these values
of $\delta $ another critical index $\eta =\frac{5-\delta }{1+\delta }$ ,
which is much better known, both experimentally and theoretically, then we
appear in the physical region with \ $\eta \approx 0.1,$ that largely (by
three times) overestimates $\eta .$ One may think that the index $\delta $
is underestimated by the Wilson $\epsilon -$expansion. So, we need by means
of the self-similar renormalization procedure to continue the asymptotic
expression (65) to the region of $\epsilon \sim 1$ with simultaneous
restoration of the incorrect limiting values at $n=-2$ and $\ n\rightarrow
\infty $. Introduce the following consecutive approximations to the quantity 
$\widetilde{\ \delta }=\delta -3:$%
\begin{equation}
\label{66} 
\begin{array}{c}
\widetilde{\delta _1}(\epsilon )=\epsilon , \\ \widetilde{\delta _2}%
(\epsilon )=\epsilon +c_2(n)\epsilon ^2. 
\end{array}
\end{equation}
By repeating the same steps that led us to (56), we obtain%
$$
\widetilde{\delta _2}^*(\epsilon ,s)=[\frac \epsilon 
{1-\frac{c_2\epsilon }{1+s}}]^{1+s}, 
$$
and 
\begin{equation}
\label{67}\delta ^{*}=3+[\frac \epsilon {1-\frac{c_2\epsilon }{1+s}}]^{1+s}. 
\end{equation}
Setting in (67) $\epsilon =1$ and $\ n=-2$ , or $n\rightarrow \infty ,$ it
is easy to show that only for $s=0$ both limits can be satisfied! Therefore 
\begin{equation}
\label{68}\delta ^{*}=3+\frac \epsilon {1-c_2(n)\epsilon }. 
\end{equation}
The expression (68) in the physical region gives the following values:%
$$
\begin{array}{c}
\delta ^{*}(n=0)=4.882,\quad \ \delta ^{*}(n=1)=4.862,\quad \\ 
\delta ^{*}(n=2)=4.852,\quad \ \delta ^{*}(n=3)=4.847. 
\end{array}
$$

The index $\eta ,$ corresponding to these values can be easily obtained from
the scaling law:%
$$
\begin{array}{c}
\eta (n=0)=0.02,\quad \eta (n=1)=0.024,\quad \\ 
\eta (n=2)=0.025,\quad \eta (n=3)=0.026. 
\end{array}
$$

These results better agree with the general understanding of $\eta $ as of a
small index and are much closer to the results of the Borel summation than
the initial $\eta =0.1,$ obtained from the $\epsilon -$expansion (65) and
the scaling law.

We conclude that the application of the self-similar renormalization
improves the Wilson $\epsilon -$expansion for the index $\delta $ both
qualitatively and quantitatively. This example stresses once again an
importance of obeying different reasonable limits in the expressions for the
critical indices. Another illustration is given in the next Section where
the self-similar renormalization of $\frac 1n-$expansion and of the $%
\epsilon -$expansion around lower critical dimension two (in powers of $d-2)$
is considered.

\section{Inverse Large Component Expansion (1/n) and Expansion in Powers of
d-2 (2+$\epsilon ).$}

The large $n-$expansions ($\frac 1n-$expansion) [18,22] and $\epsilon -$%
expansion around the lower critical dimension two ($d-2-$expansion) [23] had
raised high expectations as an alternative to the Wilson $\epsilon -$%
expansion and field-theoretical approach [2]. Nevertheless they had never
became competitive, remaining a useful guide to the region of large $n$ and
a good qualitative tool, when different aspects of the two-dimensional
behavior are considered. It is clear that the values of critical indices
given by $\frac 1n$ and $\ d-2-$expansion do not obey the $n=-2\ $ Gaussian
limit, becoming divergent at $n=0$ and $\ n=2\ (d=3),\ $respectively.
Therefore, it is not possible to get a reasonable estimate for $\nu $ and $%
\eta $ in the physical region $n=1,2,3\ $ lying too close to the spurious
pole and too far from the correct $\ n\rightarrow \infty \ $ limit,
supported by both expansions.

Consider the $\frac 1n-$expansion for the critical index $\gamma :$%
$$
\gamma =2-\frac{24}{\pi ^2n}, 
$$
from where 
$$
\gamma (n=1)=-0.432,\quad \gamma (n=2)=0.784,\quad \gamma (n=3)=1.189. 
$$
Correspondingly, the two approximations in powers of $\frac 1n$ can be
written down%
$$
\begin{array}{c}
\gamma _0(n)=2, \\ 
\gamma _1(n)=2-\frac{24}{\pi ^2n}. 
\end{array}
$$

Proceeding in accordance with the self-similar renormalization
prescriptions, the multiplier $m_1(n,s)$ can be found: 
\begin{equation}
\label{69}m_1(n,s)=1-\frac{12}{\pi ^2}\frac{s-1}{sn}. 
\end{equation}
As is seen, the minimum of $\mid m_1(n,s)\mid $ for $n\geq 2\ $is reached
for $s\rightarrow \infty .$ This gives 
$$
\gamma _1^{*}(n)=2\exp (-\frac{12}{\pi ^2n}), 
$$
and, correspondingly,%
$$
\gamma _1^{*}(n=2)=1.089,\quad \gamma _1^{*}(n=3)=1.334. 
$$

We see that for $n=3$ the value given by $\gamma _1^{*}$ becomes reasonable,
deviating from the result of the Borel-summation $\gamma =1.386\ $[20] with
the percentage error equal to $-3.752\%,$ while the initial $\frac 1n-$%
expansion has the percentage error of $-14.214\%.$ For $\ n=1,\ $the minimum
of $\mid m_1(n,s)\mid $ is reached for $s=5.633;$ correspondingly $\gamma
_1^{*}(n=1)=0.508.$ We observe that for $n=1,2$ \ the results are improved
if compared to the initial $\frac{\ 1}n-$expansion .

The $\frac 1n-$expansion for critical index $\eta $ is given as follows
[18]: 
\begin{equation}
\label{70}\eta =\frac 8{3\pi ^2n}-(\frac 83)^3\frac 1{\pi ^4n^2}. 
\end{equation}
This equation becomes singular at $n=0$ and negative at $n=-2,\ $so that,
despite its correct by design behavior at $\ n\rightarrow \infty ,$ the
values given by (70) at $n=1,2,3$ are too large:%
$$
\eta (n=1)=0.076,\quad \eta (n=2)=0.086,\quad \eta (n=3)=0.068. 
$$
The direct application of the self-similar renormalization using $\frac 1n$
\ as a renormalization parameter with%
$$
\eta _1=\frac 8{3\pi ^2}\frac 1n,\qquad \eta _2=\frac 8{3\pi ^2}\frac
1n-(\frac 83)^3\frac 1{\pi ^4}\frac 1{n^2}, 
$$
as consecutive approximations does not improve the situation, since the
influence of singularity at $n=0\ $ is too strong. To avoid this divergence,
we re-expanded the expression (70) in powers of the parameter$\ \ y=\frac{
(n+2) }{(n+8)^2},$ expressing $n$ as a function of$\ \ y:$%
\begin{equation}
\label{71}n=\frac 1{2y}(1-16y+\sqrt{1-24y}). 
\end{equation}
This choice of the re-expansion parameter is not unique, but the combination 
$\frac{n+2}{(n+8)^2}$ \ frequently appears in the Wilson $\epsilon -$%
expansion. Such a transformation restores the correct value of $\eta $ at $%
n=-2$ and also keeps intact the correct limit at $\ n\rightarrow \infty .$
Up to the third order in $y$ we obtain:%
$$
\eta =ay+by^2+cy^3,\qquad a=\frac 8{3\pi ^2},\quad b=\frac{112}{3\pi ^2}%
,\quad c=\frac{1856}{3\pi ^2}-\frac{14336}{27\pi ^4}. 
$$
Thus the following consecutive approximations may be written down:%
$$
\begin{array}{c}
\eta _1(y)=ay, \\ 
\eta _2(y)=ay+by^2, \\ 
\eta _3(y)=ay+by^2+cy^3. 
\end{array}
. 
$$

Proceeding in the usual manner, we obtain 
\begin{equation}
\label{72}\eta _2^{*}(y)=ay\exp (\frac bay), 
\end{equation}

\begin{equation}
\label{73}\eta _3^{*}(y)=(ay+by^2)\exp (\frac cay^2). 
\end{equation}
Returning to the initial variable we have%
$$
\begin{array}{c}
\eta _2^{*}(n=0)=0.013,\quad \eta _2^{*}(n=1)=0.016,\quad \\ 
\eta _2^{*}(n=2)=0.018,\quad \eta _2^{*}(n=3)=0.019, 
\end{array}
$$
and%
$$
\begin{array}{c}
\eta _3^{*}(n=0)=0.015,\quad \eta _3^{*}(n=1)=0.02,\quad \\ 
\eta _3^{*}(n=2)=0.023,\quad \eta _3^{*}(n=3)=0.025. 
\end{array}
$$

The values of multipliers in these cases are as follows:%
$$
\begin{array}{c}
m_2(y,s)=1+\frac ba 
\frac{2+s}{1+s}\ y, \\ m_3(y,s)=m_1(y,s)+\frac ca\frac{3+s}{1+s}\ y^2, 
\end{array}
$$
These values are very close to each other, e.g \ for $n=3,\; m_2=1.022$ \ and
\ $m_3=1.037.$ From the stability viewpoint the corresponding approximations
are almost equivalent. It is also possible to improve results for $\eta $ by
applying the second self-similar renormalization, as in the recurrence
relation (35). This gives $\eta ^{*}$ in the form of a continued exponential
[24]:%
$$
\eta ^{*}=ay\exp (\frac bay\exp (\frac cby)), 
$$

so that%
$$
\begin{array}{c}
\eta ^{*}(n=0)=0.016,\quad \eta ^{*}(n=1)=0.024,\quad \\ 
\eta ^{*}(n=2)=0.03,\quad \eta ^{*}(n=3)=0.032. 
\end{array}
$$

These values, especially for $n=2,3,$ are quite reasonable. The results for $%
n=0,$ not surprisingly, remain too small, since we used for the
renormalization procedure only the large $n$ expansion, obviously too short
of information about the limit for small $n.$ In order to weaken the
influence of the particular way of defining coefficients $a,b,c,$ it is
possible to proceed with a variational-optimization procedure, considering\ $%
\eta ^{*}$ as a function of two unknown parameters $\overline{a},\ \overline{%
b}$ \ and determining them from the conditions 
$$
\frac{\partial \eta ^{*}}{\partial \overline{a}}=0,\qquad \frac{\partial
\eta ^{*}}{\partial \overline{b}}=0. 
$$

For the particular choice%
$$
\eta ^{*}=\overline{a}y\exp (\frac{\overline{b}}{\overline{a}}y\frac
1{1-\frac c{\overline{b}}y}), 
$$
we find that $\ \eta ^{*}=4ecy^3$ and 
$$
\begin{array}{c}
\eta ^{*}(n=0)=0.019,\quad \eta ^{*}(n=1)=0.032,\quad \\ 
\eta ^{*}(n=2)=0.04,\quad \eta ^{*}(n=3)=0.043. 
\end{array}
$$

The $\epsilon -$expansion with $\epsilon =d-2$ around the lower critical
dimension for the critical index $\nu $$,\ $is written in the form 
\begin{equation}
\label{74}\nu ^{-1}=d-2+\frac{(d-2)^2}{n-2} 
\end{equation}
for $d>2$ and $n>2.$ At $n=3,\ d=3,\ \nu =0.5,$ giving rather crude estimate
coinciding with the mean-field result. The self-similar renormalization
using $\epsilon $ as a parameter for renormalization does not improve this
value, because $\nu \rightarrow \infty $ \ in the starting point $d=2,\ n=3.$
On the other hand, at $n\rightarrow \infty ,\ d=3,$ formula (74) gives $\nu
=1,$ i.e the correct limiting value known rigorously for the spherical
model. Re-expanding (74) in powers of $\frac 1n$ around this correct value
we obtain 
\begin{equation}
\label{75}\nu ^{-1}=1+\frac 1n+2\frac 1{n^2}+... 
\end{equation}
Proceeding in accordance with the standard prescription and using $\frac 1n$
as a renormalization parameter, we define 
$$
\begin{array}{c}
\nu _0^{-1}=1, \\ 
\nu _1^{-1}(n)=1+\frac 1n. 
\end{array}
$$

Then, we readily obtain for $\nu _1^{*}(n)$ the following expression%
$$
\nu _1^{*}(n)=\exp (-\frac 1n) 
$$
and $\ \nu _1^{*}(n=3)=0.717,$ giving reasonable estimate for the critical
index $\nu .$ The percentage error, when compared to the result of the Borel
summation $\nu =0.705$ [20] equals $1.702\%,$ while for the initial \ $d-2-$%
expansion it equals $-29.078\%.$

For the critical index $\eta ,$ the $d-2-$expansion has the following form
[23]:%
$$
\eta =a(n)(d-2)-b(n)(d-2)^2,\qquad a(n)=\frac 1{n-2},\quad b(n)=\frac{n-1}{%
(n-2)^2}. 
$$
At $n=3,\ d=3,\ $this gives $\eta =-1,\ $in disagreement with all known
about this index.

We use below $d-2=\epsilon $ as a renormalization parameter, since at $d=2,\
\eta =0,\ $being a reasonable starting point for the trajectory. Define the
following approximations to $\eta :$%
$$
\begin{array}{c}
\eta _1(\epsilon ,n)=a(n)\epsilon , \\ 
\eta _2(\epsilon ,n)=\eta _1(\epsilon ,n)-b(n)\epsilon ^2. 
\end{array}
$$

The multiplier $m_2(\epsilon ,n)=1-\frac{b(n)}{a(n)}\frac{2+s}{1+s}\epsilon$ 
reaches its minimum at $\ s\rightarrow \infty ,$ therefore%
$$
\eta _2^{*}(\epsilon ,n)=a(n)\epsilon \exp \{-\frac{b(n)}{a(n)}\epsilon \}, 
$$
and $\ \eta _2^{*}(1,3)=0.135,$ which is a considerable improvement if
compared to $\eta =-1$ from the $d-2$-expansion.

We conclude that the self-similar renormalization improves the quality of
estimates also for $\frac 1n$ and $d-2$-expansions, achieving the
quantitative agreement with other approaches. However, the broken $n=-2$
Gaussian limit still makes the possibilities of improving the results very
narrow, usually an improvement is achieved for $n=3,$ but not for lower $n.$

In the next Section we consider an opposite case of an expansion obeying
only $n=-2$ limit, but with broken limit at $n\rightarrow \infty .$ Such
situation is similar to that of Section 5, but no non-physical Gaussian
points will be present.

\section{Self-Avoiding Walk Problem (n+2-Expansion for n=0)}

Interesting properties of the LG model for $n=-2$ have been analyzed in a
number of works [14-17]. Physically, $n=-2$ corresponds to a Gaussian
polymer with the exponents $\gamma =1,\ \eta =0,\ \nu =1/2.$ From the
scaling laws in $\ d=3$ one can see that $\alpha =\frac 12,\ \delta =5,\
\beta =\frac 14.$ It seems natural to develop expansions in powers of $n+2$ (%
$n+2-$expansion) around this well defined limit [16]. To our knowledge, this
idea has never been put into practice. The task of obtaining the $n+2-$%
expansion is simplified if we note that the Wilson $\epsilon -$expansion for
the critical indices $\gamma ,\eta ,\nu ,\alpha ,\beta $ obeys the $n=-2$
limiting values. In order to obtain the $n+2-$expansion we simply re-expand
the Wilson $\epsilon -$expansion at $\epsilon =1$ in powers of$\ n+2.$ Of
course, $n+2-$expansion could be derived also from the ''first principles''
in a way similar to the $\epsilon -$expansion , or $\frac 1n-$expansion. The
nearest to the point $n=-2$ physically interesting case is located at $n=0,$
corresponding to the self-avoiding walk problem equivalent to a polymer. We
believe that the case of the order parameter dimensionality of $n=-2$ and $%
n=0$ are closely connected in a way similar to the connection existing
between the space dimensionalities $d=4$ and $d=3,$ with the only difference
that the $\epsilon -$expansion is substituted by the $n+2-$expansion. We
apply below the self-similar renormalization to the $n+2-$expansion for the
critical indices, presenting only the results for $n=0.$ The values of the
critical indices are not as good for $n=1,2,3.\ $The expansion parameter is
too large in the latter cases and also the $n\rightarrow \infty $ limit is
violated, so that the trajectory strongly deviates for larger $n\ $ from the
correct but distant starting point.

Up to the second order in $\epsilon ,$ the critical index $\nu $ is given as
follows: 
\begin{equation}
\label{76}\nu =\frac 12+\frac{n+2}{4(n+8)}\epsilon +\frac{n+2}{8(n+8)^3}%
(n^2+23n+60)\epsilon ^2. 
\end{equation}
Put here $\epsilon =1$ and expand (76) in powers of $\ z=n+2,$ up to the
third order terms in $z:$%
$$
\nu =\frac 12+a_1z+a_2z^2+a_3z^3;\qquad a_1=\frac 5{96},\quad a_2=-\frac
1{864},\quad a_3=-\frac 7{3456}. 
$$

The following approximations to the quantity $\ \widetilde{\nu }=\nu -\frac
12$ can be readily written down: 
\begin{equation}
\label{77} 
\begin{array}{c}
\widetilde{\nu _1}(z)=a_1z, \\ \widetilde{\nu _2}(z)=a_1z+a_2z^2, \\ 
\widetilde{\nu _3}(z)=a_1z+a_2z^2+a_3z^3. 
\end{array}
\end{equation}
Following the standard way, we find the multipliers 
\begin{equation}
\label{78} 
\begin{array}{c}
m_2(z,s)=1+ 
\frac{a_2}{a_1}\frac{2+s}{1+s}z, \\ m_3(z,s)=m_2(z,s)+\frac{a_3}{a_2}\frac{%
3+s}{1+s}z^2. 
\end{array}
\end{equation}
Both multipliers at the point $z=2\ (n=0),$ reach their minimum at $s=0,$
where $m_2(2,0)=0.911,\quad m_3(2,0)=0.444.$ Consequently, the trajectory
restored using all three approximations from (77) will be more stable, than
that restored from only two approximations, both trajectories being stable.
The evolution integral (12) gives%
$$
\widetilde{\nu _2}^*(z)=a_1z\frac 1{1-\frac{a_2}{a_1}z},\quad \quad 
\widetilde{\nu _3}^*(z)=\frac{\widetilde{\nu _2(z)}}{(1-2\frac{a_3}{a_1^3} 
\widetilde{\nu _2(z)}^2)^{\frac 12}}, 
$$
and $\ \nu _2^{*}(z=2)=0.6,\ \nu _3^{*}(z=2)=0.588.$ The former value is
exactly the Flory ''mean-field'' exponent [25], and the latter is the same
as the Borel-summation result, considered as the best known estimate for
polymers [20]. The latter value $\nu _3^{*}=0.588$ should be trusted more,
since it is obtained moving along the more stable trajectory than $\nu
_2^{*}.$ \ It is encouraging that the Flory and field-theory results, in our
consideration come out as successive approximations.

The Wilson $\epsilon -$expansion for the critical index $\beta ,\ $up to the
second order in $\epsilon ,\ $is 
\begin{equation}
\label{79}\beta =\frac 12-\frac 3{2(n+8)}\epsilon +\frac{(n+2)(2n+1)}{%
2(n+8)^3}\epsilon ^2. 
\end{equation}
So, the $n+2-$expansion, up to the third order, becomes%
$$
\beta =a_0+a_1z+a_2z^2+a_3z^3;\qquad a_0=\frac 14,\quad a_1=\frac
5{144},\quad a_2=\frac 1{864},\quad a_3=-\frac 1{432}. 
$$

This results in the following approximations for $\beta :$%
\begin{equation}
\label{80} 
\begin{array}{c}
\beta _0(z)=a_0, \\ 
\beta _1(z)=\beta _0(z)+a_1z, \\ 
\beta _2(z)=\beta _1(z)+a_2z^2, \\ 
\beta _3(z)=\beta _2(z)+a_3z^3. 
\end{array}
\end{equation}
The multipliers $m_1(z,s),\ m_2(z,s),\ m_3(z,s)$ \ for $\ z=2$ reach their
minima at $\ s\rightarrow \infty ;$ so that $\ m_3(2,\infty )$ $<$ $%
m_2(2,\infty )<$ $m_1(2,\infty ).$ The evolution-integral solution for $%
\beta _3^{*}(z)$ is 
\begin{equation}
\label{81}\beta _3^{*}(z)=\beta _2(z)\exp (\frac{a_3}{a_0}z^3), 
\end{equation}
and $\ \beta _3^{*}(z=2)=0.301.\ $ This coincides with the result of the
Borel summation [20].

The critical index $\eta $ has the following $\epsilon -$expansion, up to
the third order: 
\begin{equation}
\label{82}\eta =\frac{n+2}{2(n+8)^2}\epsilon ^2+\frac{n+2}{8(n+8)^4}%
(272+56n-n^2)\epsilon ^3, 
\end{equation}
and the corresponding $n+2-$expansion, up to the third order terms, can be
obtained :%
$$
\eta =a_1z+a_2z^2+a_3z^3;\qquad a_1=\frac{25}{864},\quad a_2=-\frac{23}{2592}%
,\quad a_3=\frac{43}{31104}. 
$$

Thus, the following approximations result%
$$
\begin{array}{c}
\eta _1(z)=a_1z, \\ 
\eta _2(z)=a_1z+a_2z^2, \\ 
\eta _3(z)=a_1z+a_2z^2+a_3z^3.
\end{array}
$$
The multipliers $m_2(z,s)$ and $\ m_3(z,s)$ at $z=2$ satisfy the condition $%
\mid m_2(z,s)\mid <1,\mid m_3(z,s)\mid <1\ $for arbitrary $s,\ $the former
satisfying the condition $\mid m_2(z,s)\mid =0$ at $s=0.586,$ the latter
becoming minimal at $s=0.\ $Thus, we find%
$$
\eta _2^{*}(z)=a_1z\frac 1{(1-\frac{a_2}{a_1(1+s)}z)^{1+s}},\quad \quad \eta
_3^{*}(z)=\frac{\eta _2(z)}{(1-2\frac{a_3}{a_1^3}\eta _2(z)^2)^{\frac 12}}, 
$$
with $\ \eta _2^{*}(z=2)=0.034,\ \ \eta _3^{*}(z=2)=0.023.$ The former value
is very close to the so-called unconstrained $\epsilon -$expansion ($\eta
=0.031\pm 3\ $[26]) and the constrained $\epsilon -$expansion ($\eta
=0.0320\pm 25\ $[26]). The latter value approaches closely the result of
Borel summation, $\eta =0.027\pm 4\ $[26]. The scaling law $\frac 12\nu
(1+\eta )=\beta $\ \ is ideally satisfied with our $\nu _3^{*}=0.588,\ \eta
_3^{*}=0.023$ and $\ \beta _3^{*}=0.301.$

It is also worth noting that a non-optimal, but stable trajectory \ for $%
\eta _3^{*}(z,s\rightarrow \infty )=\eta _2(z)\exp (\frac{a_3}{a_1}z)$ leads
us to the point $\eta _3^{*}(2,\infty )=0.027,\ $which is exactly the value
of the Borel summation.

The $\epsilon -$expansion for the index $\gamma ,$ up to the second order,
has the form 
\begin{equation}
\label{83}\gamma =1+\frac{n+2}{2(n+8)}\epsilon +\frac{n+2}{4(n+8)^3}%
(n^2+22n+52)\epsilon ^2, 
\end{equation}
and the $n+2-$expansion, up to the third order, is%
$$
\gamma =1+a_1z+a_3z^3;\qquad a_1=\frac 7{72},\quad a_3=-\frac 1{216}. 
$$

The multiplier at $z=2$ acquires its minimum value at $s=0$ and%
$$
\gamma _3^{*}(z)=\frac{a_1z}{(1-\frac{2a_3}{a_1}z^2)^{1/2}}+1 
$$
with $\gamma _3^{*}(z=2)=1.165,$ and with the percentage error $\epsilon
_3^{*}(\gamma )=0.345\%$ as compared to the results of Borel summation [26].
Moving along a non-optimal but stable trajectory with $s\rightarrow \infty ,$
we come to $\gamma _3^{*}(z,s\rightarrow \infty )=a_1z\exp (\frac{a_3}{a_1}%
z^2)+1,\ $which yields the value $\gamma _3^{*}(2,s\rightarrow \infty
)=1.161,$ in the complete agreement with the Borel summation [20,26]. The
scaling law $\ \nu (2-\eta )=\gamma \ $ gives for $\nu =\nu _3^{*},\ \eta
=\eta _3^{*}$ the value $\gamma =1.162.$ The percentage error in this case
equals $0.258\%.$

For the critical index $\alpha ,$ the $\epsilon -$expansion, up to the
second order, reads: 
\begin{equation}
\label{84}\alpha =\frac{4-n}{2(n+8)}\epsilon -\frac{(n+2)^2}{4(n+8)^3}%
(n+28)\epsilon ^2, 
\end{equation}
and the $n+2-$expansion, up to the third order, follows:%
$$
\alpha =\frac 12-a_1z-a_2z^2-a_3z^3,\quad \quad a_1=\frac 16,\quad a_2=\frac
1{432},\quad a_3=-\frac 1{108}. 
$$

For the quantity $\widetilde{\ \alpha }=\frac 12-\alpha ,$ with the set of
approximations 
$$
\begin{array}{c}
\widetilde{\alpha _1}(z)=a_1z, \\ \widetilde{\alpha _2}(z)=\widetilde{\alpha
_1}(z)+a_2z^2, \\ \widetilde{\alpha _3}(z)=\widetilde{\alpha _2}(z)+a_3z^3, 
\end{array}
$$
following the conventional prescriptions, we find the solution corresponding
to the most stable trajectory with $s=0:$%
$$
\widetilde{\alpha _3}^*(z)=\frac{\widetilde{\alpha 
_2}(z)}
{[1-2(a_3/a_1^3)\widetilde{\alpha _2}(z)]^{1/2}}, 
$$
and $\alpha _3^{*}(z=2)=0.217.$ The field-theory Borel summation results are
not available. From the scaling law $\nu =\frac{(2-\alpha)}{3},$ we find that
the value $\nu =0.594,$ corresponding to $\alpha _3^{*},$ within the
percentage error of $1\%$ agrees with the Borel summation and our own
estimate for $\nu .$ Therefore, our estimate for $\alpha $ may be considered
as satisfactory.

The critical index $\delta ,$ with the $n=-2$ \ limit violated by the Wilson 
$\epsilon -$expansion was estimated in Section 6. We found that $\delta
(n=0)=4.882.$ Theoretical field Borel summation data are not available for
comparison. From our estimate $\eta _3^{*}=0.023$ and the scaling relation $%
\delta =\frac{5-\eta }{1+\eta },$ we obtain $\delta =4.865,$ and within the
percentage error of $0.349\%$ both our estimates agree.

We believe that both good and reliable estimates can be obtained only from
expansions possessing correct limits by design. Examples of such behavior
are given by the $\epsilon -$expansion and field theory expansion in powers
of an interaction constant.

\section{Application to $\epsilon -$Expansion}

In accordance with all our previous remarks, concerning the importance of
the correct limits at $n=-2\ $and $n\rightarrow \infty ,$ we considered the
well-known Wilson $\epsilon -$expansion [1] and found that it does not obey
these limits for the critical indices $\delta ,\gamma ,\alpha $ and $\nu .$

We have seen in Section 6, that for the critical index $\delta $ both limits
are violated with a percentage error equal to $-10\%.$ The critical index $%
\gamma $ (see (83)) does obey the $n=-2$ limit, but at $\ n\rightarrow
\infty ,\ \gamma =1.75,$ and the percentage error equals $-12.5\%,$ when
compared with the exact $\gamma =2.$ The critical index $\alpha $ (see (84),
obeys the $n=-2$ \ limit, but at $\ n\rightarrow \infty ,\ \alpha =-0.75,\ $%
instead of the exact $\alpha =-1,\ $with the percentage error $-25\%.$ The
critical index $\nu $ (see (76)) obeys the $n=-2$ \ limit, but at $\
n\rightarrow \infty ,\ \ \nu =0.875,\ $with the percentage error $-14.286\%$
when compared to the exact $\nu =1.$ Clearly these discrepancies should lead
to an uncontrolled error within the physical region $n=0,1,2,3$.

Fortunately, the last two indices, $\eta \ $and $\beta $ (see (82) and (79))
do obey the $n=-2\ $and $n\rightarrow \infty $ limits: $\eta =0$ at $\ n$ $%
=-2,$ $n\rightarrow \infty $ and $\ \beta =\frac 14$ at $\ n=-2,$ and $\
\beta =\frac 12$ at $\ n\rightarrow \infty .$

Compare, in the physical region, the values of $\delta ,\gamma ,\alpha $ and 
$\nu $ obtained from $\eta \ $and $\beta $ by means of the scaling laws with
those obtained by the direct use of the Wilson $\epsilon -$expansion. The
percentage deviation from the initial Wilson expansion, for the index $%
\delta $ \ is (in this Section, in order not to cause confusion, we use the
word ''error'' instead of the letter ''$\epsilon $'' in the formula (59)):%
$$
\begin{array}{c}
E(\delta ,n=0)=7.7\%,\qquad E(\delta ,n=1)=7.21\%, \\ 
E(\delta ,n=2)=7.06\%,\qquad E(\delta ,n=3)=7.082\% 
\end{array}
. 
$$

For the index $\gamma :$%
$$
\begin{array}{c}
E(\gamma ,n=0)=0.7\%,\qquad E(\gamma ,n=1)=1.97\%, \\ 
E(\gamma ,n=2)=3.27\%,\qquad E(\gamma ,n=3)=4.49\%,
\end{array}
$$
the error increases considerably with n, while at $n=0$ the correct limit $%
n=-2$ still favorably influences the results. For the index $\alpha $ we
have 
$$
\begin{array}{c}
E(\alpha ,n=0)=-0.214\%,\qquad E(\alpha ,n=1)=-20.36\%, \\ 
E(\alpha ,n=2)=168.95\%,\qquad E(\alpha ,n=3)=52.19\%,
\end{array}
$$
again being negligible for $n=0$ but growing with $n.$ For the index $\nu $
we get%
$$
\begin{array}{c}
E(\nu ,n=0)=1.67\%,\qquad E(\nu ,n=1)=3.13\%, \\ 
E(\nu ,n=2)=4.52\%,\qquad E(\nu ,n=3)=5.75\%.
\end{array}
$$

We see that only at $n=0$ the results possess a reasonable accuracy, and the
quality of the Wilson $\epsilon -$expansion is good enough to reach, e.g. by
means of $n+2-$expansion (see Section 8), the quality of the best known
estimates.

We conclude that all attempts to improve the critical indices $\delta
,\gamma ,\alpha ,$ $\nu $ for $n=1,2,3$ directly from the Wilson $\epsilon - 
$expansion will leave us with an uncontrollable error. It seems reasonable
to renormalize self-similarly only the critical indices $\eta $ and $\beta $
possessing correct limiting values and to calculate all other indices form
the scaling laws.

For the index $\eta $ (see (82)), using $\epsilon $ as a parameter for
renormalization, the following approximations are available%
$$
\begin{array}{c}
\eta _2(\epsilon )=a_2(n)\epsilon ^2,\quad \qquad a_2(n)= 
\frac{n+2}{2(n+8)^2}, \\ \eta _3(\epsilon )=\eta _2(\epsilon
)+a_3(n)\epsilon ^3,\quad a_3(n)=\frac{n+2}{8(n+8)^4}(272+56n-n^2). 
\end{array}
$$

The multiplier $m_3(n,s)$ at $\ \epsilon =1$ is always minimal at $\
s\rightarrow \infty ,$ therefore 
\begin{equation}
\label{85}\eta _3^{*}(\epsilon )=a_2(n)\epsilon ^2\exp \{\frac{a_3(n)}{a_2(n)%
}\epsilon \ t^{*}\}. 
\end{equation}
For $t^{*}=1$ we obtain%
$$
\begin{array}{c}
\eta _3^{*}(n=0)=0.045,\quad \eta _3^{*}(n=1)=0.051,\quad \\ 
\eta _3^{*}(n=2)=0.052,\quad \eta _3^{*}(n=3)=0.05. 
\end{array}
$$

These values are probably too large. The situation may be improved if (85)
is optimized using the knowledge of the exact\ $\eta =0.2083$ \ for $%
\epsilon =2,\ n=0$ ($2$-$d$ random walks) [27].

Setting $\eta _3^{*}(\epsilon =2,n=0,t^{*})=0.2083\ $ we find that $\
t^{*}=0.567$ \ and 
$$
\begin{array}{c}
\eta _3^{*}(n=0)=0.029,\quad \eta _3^{*}(n=1)=0.033,\quad \\ 
\eta _3^{*}(n=2)=0.034,\quad \eta _3^{*}(n=3)=0.034, 
\end{array}
$$
in a perfect agreement with the best estimates by the Borel summation [20].
If $\ t^{*}$ is optimized from the knowledge of the $2d$ Ising model exact $%
\eta =0.25,$ then similarly, \ $t^{*}=0.603$ and 
$$
\begin{array}{c}
\eta _3^{*}(n=0)=0.03,\quad \eta _3^{*}(n=1)=0.034,\quad \\ 
\eta _3^{*}(n=2)=0.035,\quad \eta _3^{*}(n=3)=0.035. 
\end{array}
$$

There is also another way to get information from the $\ \epsilon -$%
expansion for the index $\eta .$ Proceed similarly to the case of $\frac 1n-$%
expansion, and re-expand (82) in powers of \ $y=\frac{n+2}{(n+8)^2}\ ($at $%
\epsilon =1)$. Then, up to the second order in $y,$%
$$
\eta \approx \frac 38y+9y^2 
$$
and $\ \eta _2^{*}=\frac 38y\exp (24y)$, \ with%
$$
\begin{array}{c}
\eta _2^{*}(n=0)=0.025,\quad \eta _2^{*}(n=1)=0.034,\quad \\ 
\eta _2^{*}(n=2)=0.039,\quad \eta _2^{*}(n=3)=0.042, 
\end{array}
$$
still in a good agreement with a set of data available for the critical
index $\eta $ [21].

For the critical index $\beta ,$ Eq.(79) defines the following set of
approximations:%
$$
\begin{array}{c}
\beta _0(\epsilon )=a_0,\qquad a_0=1, \\ 
\beta _1(\epsilon )=\beta _0(\epsilon )+a_1(n)\epsilon ,\qquad a_1(n)=-\frac
3{2(n+8)}, \\ 
\beta _2(\epsilon )=\beta _1(\epsilon )+a_2(n)\epsilon ^2,\qquad a_2(n)= 
\frac{(n+2)(2n+1)}{2(n+8)^3}. 
\end{array}
$$
The multiplier 
$$
m_2(n,s)=1+\frac{a_1(n)}{a_0}\frac{1+s}s\epsilon +\frac{a_2(n)}{a_0}\frac{2+s%
}s\epsilon ^2 
$$
is equal to zero at $s=-\frac{2a_2+a_1}{a_2+a_1+a_0},\ $and the evolution
integral gives 
\begin{equation}
\label{86}\beta _2^{*}(\epsilon )=\frac{\beta _1(\epsilon )}{\epsilon ^s}%
\frac 1{[1-\frac{2a_2^2}sa_0^{-(1+\frac s2)}\beta _1^{\frac 2s}(\epsilon
)]^{\frac s2}}. 
\end{equation}
From here, at $\epsilon =1,$%
$$
\begin{array}{c}
\beta _2^{*}(n=0)=0.313,\quad \beta _2^{*}(n=1)=0.333,\quad \\ 
\beta _2^{*}(n=2)=0.35,\quad \beta _2^{*}(n=3)=0.364. 
\end{array}
$$
The formula (86) is applicable only up to $n=42,$ where $s\rightarrow 0$ and
(86) becomes undefined. For $n\geq 42,$ the exponential summation is optimal:%
$$
\beta _2^{*}(\epsilon )=\beta _1(\epsilon )\exp (\frac{a_2(n)}{a_0}\epsilon
),\quad n\geq 42. 
$$
For $n=2,3,$ $\ \beta _2^{*}$ almost coincides with the Borel summation
values $0.346\pm 2$ and $0.365\pm 2,$ respectively, being larger for $n=0,1,$
where the Borel summation gives $0.302\pm 15,$ and $0.325\pm 1$ respectively.

The $\epsilon -$expansion was obtained also from the theoretical field
approach up to the fifth order in $\epsilon $ [28]. For the critical index $%
\nu ,$ up to the second order in $\epsilon ,$ one has 
\begin{equation}
\label{87} 
\begin{array}{c}
V\equiv \nu ^{-1}=a_a(n)+a_1(n)\epsilon +a_2(n)\epsilon ^2, \\ 
a_a(n)=2,\quad a_1(n)=-\frac{n+2}{n+8},\ \quad a_2(n)=-\frac{(n+2)(13n+44)}{%
2(n+8)^3}, 
\end{array}
\end{equation}
the limits $n=-2$ and $\ n\rightarrow \infty $ being satisfied, in
distinction from the original $\epsilon -$expansion. We use the following
approximations to the quantity $\ \widetilde{V}=-V+a_0(n):$%
\begin{equation}
\label{88} 
\begin{array}{c}
\widetilde{V}_1(\epsilon )=-a_1(n)\epsilon , \\ \widetilde{V}_2(\epsilon
)=-a_1(n)\epsilon -a_2(n)\epsilon ^2. 
\end{array}
\end{equation}
All terms in (88) are positive and, the optimal renormalization corresponds
to $s\rightarrow \infty .$ The evolution integral can be readily calculated
giving 
\begin{equation}
\label{89}\widetilde{V}_2^*(\epsilon )=-a_1(n)\epsilon \exp 
\{\frac{a_2(n) }{a_1(n)}\epsilon \ t^{*}\}. 
\end{equation}
We found that at $\ t^{*}=1,\ \epsilon =1,$%
$$
\begin{array}{c}
\nu _2^{*}(n=0)=0.607,\quad \nu _2^{*}(n=1)=0.655,\quad \\ 
\nu _2^{*}(n=2)=0.698,\quad \nu _2^{*}(n=3)=0.736. 
\end{array}
$$

Similarly to the case of the index $\eta ,$ the self-similarly renormalized $%
\epsilon -$expansion for $\nu \ $overestimates the critical index, as may be
seen from the comparison with other results [21]. Let us optimize the
expression for $\nu _2^{*}$ using the knowledge of the $2d$ Ising $\nu =1$
[29]. From the condition $\nu _2^{*}(\epsilon =2,n=1,t^{*})=1$ we find $\
t^{*}=0.576$ and the optimized values%
$$
\begin{array}{c}
\nu _2^{*}(n=0)=0.59,\quad \nu _2^{*}(n=1)=0.628,\quad \\ 
\nu _2^{*}(n=2)=0.662,\quad \nu _2^{*}(n=3)=0.691. 
\end{array}
$$

Compared to the best known calculations of $\nu $ from the Borel summation
and similar methods, for $n=0,1$ our estimates practically coincide with
them, and for $n=2,3$ \ the percentage errors are$\ -0.451\%$ compared to $%
\nu =0.665$ at $n=2$\ [30] and $1.003\%$ compared to $\nu =0.698$ at $n=3\ $
[30].

It is interesting that by a single parameter optimization the index $\nu $
is improved in the whole physical range. It is worth noting that $\ t^{*}$
used for optimization is about the same for $\eta $ and $\nu .$ Evaluation
of the critical indices is also possible basing on the field-theoretical
expansions. Information is obtained in the latter case from the perturbative
series directly and the results are marginally sensitive to the way in which
the position of an infrared stable fixed point is determined [20,31].

\section{Application to the Field Theory Expansion}

Theoretical field approach in the theory of critical phenomena gives,
probably, the most accurate and consistent estimates for the critical
indices $\eta $ and $\gamma $ [2]. The analysis of the expansions in powers
of the interaction constant $g$ ($g$-expansion) for these indices from the
viewpoint of the limiting cases $n=-2,\ n\rightarrow \infty ,$ became
possible only when the $g$-expansions have been written for arbitrary $n$
[30]. By direct inspection of the expressions for $\eta $ and $\gamma $ from
[30], we found that\ $n\rightarrow \infty $ limit is obeyed rigorously if $\
g=1,\ $i.e $\ \eta =0,\ \gamma =2$ and the $n=-2$ limit is obeyed with very
high accuracy for arbitrary $g,$ i.e $\eta \approx 0$ $\gamma \approx 1$
with the error insignificant within the framework of the theory of critical
indices.

The standard approach [20] uses, for computing the infrared stable fixed
point $g^{*}$ of the beta-function $W(g),$ a complicated Borel summation
technique. Then critical indices are calculated as $\gamma (g^{*}),\ \eta
(g^{*}).\ $This approach requires a number of terms in the expansions. We
apply below the self-similar renormalization to only the initial three terms
in the expressions for $\ W(g),\gamma ^{-1}(g),\ \eta (g)$ \ and obtain
estimates with an accuracy comparable to the best known Borel summation
results obtained from all known terms in the expansions.

We construct the following set of approximations to $W(g)$ using the $g-$%
expansion from [30]: 
\begin{equation}
\label{90} 
\begin{array}{c}
W_2(g)=-g+g^2, \\ 
W_3(g)=W_2(g)-a_3(n)\ g^3,\qquad a_3(n)=\frac{6.07407408\ n+28.14814815}{%
(n+8)^2}. 
\end{array}
\end{equation}
From the equation $W_2(g)=f$ \ we find the expansion function $x(f)=\frac
12(1+\sqrt{1+4f}$ ). The points of the approximation cascade trajectory are%
$$
\begin{array}{c}
y_2(f)=f, \\ 
y_3(f)=y_2(f)-\frac{a_3}8x^3(f). 
\end{array}
$$
For the velocity field we get 
$$
\text{{\it v}}_3(f)=-\frac{a_3}8x^3(f). 
$$
By analogy with Section 4, the evolution-integral solution for $W_3^{*}$ is
obtained implicitly from the equation 
$$
\int_{W_2(g)}^{W_3^{*}}\frac{df}{\text{{\it v}}_3(f)}=t^{*}, 
$$
and the root $\ g^{*}$ of the equation $\ W_3^{*}(g,n,t^{*}=1)=0$ \ is
obtained numerically, as a function $\ g^{*}=g^{*}(n).$ In the physically
important cases%
$$
\begin{array}{c}
g^{*}(n=0)=1.59,\quad g^{*}(n=1)=1.559,\quad \\ 
g^{*}(n=2)=1.524,\quad g^{*}(n=3)=1.491. 
\end{array}
$$

At $\ n=-2,\ g^{*}=1.599$ and at $\ n\rightarrow \infty ,\ g^{*}=1.$ The
dependence of $g^{*}(n)$ in the interval $n=-2,0$ is nonmonotonous, a
maximum is reached at $n=-1,$ where $g^{*}=1.61.\ $Our values are higher
than the results of [30], but remain within the reasonable bounds and show
the same tendency, at least for $n=0,\infty .$ No data are available for
comparison for $n=-2,0.$

For the critical index $\eta ,$ keeping the starting two terms in powers of $%
g,$ we can write down the following approximations 
\begin{equation}
\label{91} 
\begin{array}{c}
\eta _2(g)=b_2(n)g^2,\qquad b_2(n)= 
\frac{0.2962962963(n+2)}{(n+8)^2}, \\ \eta _3(g)=\eta _2(g)+b_3(n)g^3,\quad
b_3(n)=\frac{0.0246840014n^2+0.246840014n+0.3949440224}{(n+8)^3}, 
\end{array}
\end{equation}
and the evolution integral can be readily calculated resulting in%
$$
\eta _3^{*}(g,n)=b_2(n)g^2\exp \{\frac{b_3(n)}{b_2(n)}g\}, 
$$
and%
$$
\begin{array}{c}
\eta _3^{*}(n=0)=0.027,\quad \eta _3^{*}(n=1)=0.030,\quad \\ 
\eta _3^{*}(n=2)=0.031,\quad \eta _3^{*}(n=3)=0.031. 
\end{array}
$$
These values are practically the same that quoted in [20] where the Borel
summation has been used: $\eta (n=0)=0.026\pm 3,\ \eta (n=1)=0.031\pm 4,\
\eta (n=2)=0.033\pm 4,\ \eta (n=3)=0.033\pm 4.\ $At $n=4,\ \eta _3^{*}\ $%
decreases to $0.03,$ the same tendency is seen in the data of [30].

For the critical index $\gamma $ we keep the three starting terms in powers
of $g:$%
\begin{equation}
\label{92}G\equiv \gamma ^{-1}=1+c_1(n)g-c_2(n)g^2,\qquad c_1(n)=-\frac{n+2}{%
2(n+8)},\quad c_2(n)=-\frac{n+2}{(n+8)^2}. 
\end{equation}
The following approximations to the quantity $\widetilde{G}$ $=-G+1$ are
used:%
$$
\begin{array}{c}
\widetilde{G}_1(g)=c_1(n)g, \\ \widetilde{G}_2(g)=\widetilde{G}%
_1(g)-c_2(n)g^2. 
\end{array}
$$
The multiplier $m_2(g,n,s)=1-\frac{c_2(n)}{c_1(n)}\frac{2+s}{1+s}g$,
calculated for $g=g^{*}(n),$ is always minimal for $s=0.\ $The evolution
integral gives 
$$
\widetilde{G}_2^*(g,n)=c_1(n)g\frac 1{1+\frac{c_2(n)}{c_1(n)}g}. 
$$

For $\gamma _2^{*}$ we obtain 
$$
\begin{array}{c}
\gamma _2^{*}(n=0)=1.166,\quad \gamma _2^{*}(n=1)=1.239,\quad \\ 
\gamma _2^{*}(n=2)=1.305,\quad \gamma _2^{*}(n=3)=1.363. 
\end{array}
$$

The percentage errors are $\epsilon _2(\gamma (n=0))=0.431\%,$ as compared
to $1.161$ from the Borel summation: \ zero error as compared to $1.241\pm 2$
at $\ n=1;\ \epsilon _2(\gamma (n=2))=-0.836\%$ as compared to\ 1.316, and $%
\ \epsilon _2(\gamma (n=3))=-1.659\%,$ as compared to $1.386$ from the Borel
summation. In the latter case of $n=3$ we also constructed the set of
approximations directly for the index $\gamma ,$ expanding (92) in powers of 
$g,$ up to the third order : 
\begin{equation}
\label{93}\gamma =1+d_1(n)g+d_2(n)g^2,\qquad d_1(n)=-c_1(n),\quad
d_2(n)=c_1^2(n)+c_2(n). 
\end{equation}
Approximating $\widetilde{\gamma }$ $=\gamma -1$ by%
$$
\begin{array}{c}
\widetilde{\gamma _1}(n)=d_1(n)g, \\ \widetilde{\gamma _2}(n)=\widetilde{%
\gamma _1}(n)+d_2(n)g^2, 
\end{array}
$$
and finding out that the multiplier $m_2(g,n,s)\ \ $is minimal at $%
s\rightarrow \infty ,$ we have 
\begin{equation}
\label{94}\widetilde{\gamma _2}^*(n)=d_1(n)g\exp 
\{\frac{d_2(n)}{d_1(n)}g\ t^{*}\}. 
\end{equation}

At $\ t^{*}=1$ we obtain $\gamma _2^{*}(n=3)=1.37$ and the percentage error $%
-1.154\%.$ The same procedure applied to $n=0,1,2$ always keeps the error $%
\epsilon <1\%.\ $These results are quite accurate, especially if to remember
that only the starting terms were used. Unfortunately, the expression (93)
does not obey the $n\rightarrow \infty $ limit with a percentage error $%
-3.7\%.$ An optimization of the expression (94) may be carried out requiring
that the limit $\gamma =2$ at $\ n\rightarrow \infty $ be restored (see also
section 5). This aim is achieved with $t^{*}=1.15\ \ $and $\gamma
_2^{*}(n=3)=1.375$ with a percentage error $-0.794\%.$ An effective increase
of $\ t^{*}$ mimics the effect of \ introducing higher terms into
consideration.

\section{Conclusion}

We suggested here a new variant in the method of self-similar
renormalization permitting to find effective sums of asymptotic series. The
advantage of the method is that it allows to get results exploiting just a
few first terms of given series, when no other resummation techniques work.
At the same time, the accuracy of the results is not worse than that reached
in other known sophisticated techniques involving about ten terms. In
addition, our method in the majority of cases, makes it possible to present
answers in the form of simple analytical expressions that are easy to study
for considering their dependencies on various parameters, including
asymptotic behavior with respect to these parameters.

The possibility of realizing a renormalization, having in hands only a few
terms of a series, is due to an algebraic transformation which is equivalent
to the effective increase of approximation orders. That is why we call this
variant the algebraic self-similar renormalization.

The general idea of the self-similar renormalization [4-8] is to extract the
maximum of information from the minimal number of terms. Such a minimax
criterion, certainly, can be followed only with the help of additional
functions making the convergence as fast as possible. These functions are
called control or governing functions, and they play the same role as the
control functions in the optimal control theory. In the algebraic
self-similar renormalization, the control functions are introduced into
powers of an algebraic transformation. The choice of these control functions
is based on the principle of maximal stability by minimizing the absolute
value of mapping multipliers.

We illustrated the effectiveness of our approach by renormalizing divergent
series in the theory of critical phenomena. Doing this, we specially
restricted ourselves from using many terms of perturbative series, which are
sometimes known--This is to emphasize that our approach is effective when,
really, only a minimal information is available. Dealing with higher-order
terms needs a multiple repetition of the renormalization procedure. This
requires a slight generalization of the technique and much more place for
presentation. This multiple renormalization will be the subject of a
separate publication.

The method suggested is quite general and can be applied to arbitrary
divergent series. The choice of examples from the theory of critical
phenomena owes to the common interest to this problem. We wanted also to
stress that even for the problem, where so much is known, there are ways of
improvement by first, obtaining the results much easier, second, deriving
analytical formulas, not involving heavy numerical calculations, and,
finally, by restoring correct asymptotic behavior with respect to physical
parameters, such as the number of components.

Another important message which we would like to bring up to readers is that
one should not be afraid of simple perturbative series that, being
divergent, seem, at the first glance, to be senseless: Even a seemingly bad
perturbative series contains quite a lot of useful information which can be
extracted by means of an efficient renormalization procedure. We hope we
were able to convince the reader that the algebraic self-similar
renormalization suggested can be such a tool for extracting quite accurate
information even from bad and short series.

This work was supported by a grant from the National Science and Technology
Devepolement Council of Brazil.

\end{document}